\documentclass[letter]{aa} % for the letters 
\usepackage{graphicx}
\usepackage{dblfloatfix}
\usepackage{txfonts}
\begin{document} 
   \title{Measuring the atomic composition of planetary building blocks.}
   %\subtitle{}

   \author{M. K. McClure\inst{1, 2}, C. Dominik\inst{1}, M. Kama\inst{3, 4}}

\institute{$^1$ Anton Pannekoek Institute for Astronomy,
                         Universiteit van Amsterdam,
                         Science Park 904, 
                         1098 XH Amsterdam, Netherlands \\
            $^2$ Leiden Observatory, 
   			Leiden University, 
   			PO Box 9513, 
   			2300 RA Leiden, The Netherlands \\
            $^3$ Department of Physics and Astronomy, University College London, Gower Street, London, WC1E 6BT, UK \\
            $^4$ Tartu Observatory, University of Tartu,
                            Observatooriumi 1, T\~{o}ravere, 61602, Estonia \\
              \email{mcclure@strw.leidenuniv.nl}
                   }

   \date{Received -; accepted -}

% \abstract{}{}{}{}{} 
% 5 {} token are mandatory

  \abstract
   {Volatile molecules are critical to terrestrial planetary habitability, yet difficult to observe directly where planets form at the midplanes  of  protoplanetary  disks.  It  is  unclear  whether  the  inner $\sim$1  AU  of  disks  are  volatile-poor or if this region is resupplied with ice-rich dust from colder disk regions. Dust traps at radial pressure maxima bounding disk gaps can cut off the inner disk from such volatile reservoirs. However, the trap retention efficiency and atomic composition of trapped dust have not been measured.}
  %It is unclear whether the inner $\sim$1 AU of disks are volatile-poor, leading to Earth-like extrasolar planets, or if this region is resupplied with ice-rich dust from colder disk regions. 
   {We present a new technique to measure the absolute atomic abundances in gas accreting onto T Tauri stars and infer the bulk atomic composition and distribution of midplane solids that have been retained in the disk around the young star TW Hya.}
   {We identify near-infrared atomic line emission from gas-phase material inside the dust sublimation rim of TW Hya. Gaussian decomposition of the strongest H Paschen lines isolates the inner disk hydrogen emission. We measure several key elemental abundances, relative to hydrogen, using a chemical photoionization model and infer dust retention in the disk. With a 1D transport model, we determine approximate radial locations and retention efficiencies of dust traps for different elements.}
   {Volatile and refractory elements are depleted from TW Hya's hot gas by factors of $\sim$10$^2$ and up to 10$^{5}$, respectively. The abundances of the trapped solids are consistent with a combination of primitive Solar system bodies. Dust traps beyond the CO and N$_2$ snowline cumulatively sequester 96\% of the total dust flux, while the trap at 2 AU, near the H$_2$O snowline, retains 3\%. The high depletions of Si, Mg, and Ca are explained by a third trap at 0.3 AU with $>$95\% dust retention.}
  {TW Hya sports a significant volatile reservoir rich in C- and N-ices in its outer sub-mm ring structure. However, unless the inner disk was enhanced in C by earlier radial transport, typical C destruction mechanisms and lack of C resupply should leave the terrestrial planet-forming region of TW Hya 'dry' and carbon-poor. Any planets that form within the silicate dust trap at 0.3 AU could resemble Earth in terms of the degree of their volatile depletion.}
   
   %effectively blocking replenishment of the inner disk solids from cold reservoirs. 
   %Volatile-enhancement of giant planet atmospheres cannot be explained by primary accretion of this gas. 


   \maketitle
%
%--------------------------------------------------------------------------------------------------------------------------------------------------------------------------------------------------------------------------------------------------------
%--------------------------------------------------------------------------------------------------------------------------------------------------------------------------------------------------------------------------------------------------------
%--------------------------------------------------------------------------------------------------------------------------------------------------------------------------------------------------------------------------------------------------------

\section{Introduction}
Relative abundances of common elements, e.g. C, O, and Si, are used to characterize both planetary atmospheric compositions \citep{madhusudhan12b} and those of solar system solid bodies \citep{bergin15}. Despite their high solar abundances and enhancement in gas giant atmospheres, C, O, and N are depleted relative to Si on Earth and all classes of meteorites \citep{bergin15, lodders2010}. The fraction of each element in rocks (`refractories') versus ices (`volatiles') may influence these trends. Ice sublimation depletes solids, and additional mechanisms may exist to destroy C-rich rocky grains in the inner regions of protoplanetary disks (PPDs) while retaining those rich in Si \citep{anderson17}. Radial transport of fresh solids from the cold outer disk could renew the C in the terrestrial planet forming region of PPDs, and it may be that the low C abundances in solar system bodies are a consequence of Jupiter segregating the inner solar PPD, rather than of destruction by a universal mechanism \citep{klarmann18}. However, the midplane distribution of volatile-rich solids in protoplanetary disks is not yet observationally well-determined, due to high optical depths. Moreover, it is unclear where or when these solids start being retained in the disk, or `locked' out of the gas accreting onto the central star, a key step in the formation of planets \citep{drake2005}.

The presence of midplane ice reservoirs in the nearby disk TW Hya has been inferred from cold gas depletions of CO and water by factors of $\sim$100 and $\sim$800, respectively, relative to the values typically measured for the interstellar medium (ISM) \citep{schwarz+16a, hogerheijde+11}, depending on the disk chemistry \citep{kamp2013}. These missing gas species are thought to be located beyond `snowlines', the locations in PPDs where the midplane temperature is low enough for gas of a given species to freeze out onto dust grains, e.g 4 AU for water and 20 AU for CO in TW Hya \citep{zhang13, zhang2017}. If icy grains grow to at least millimeter sizes, then they no longer vertically circulate into the region where the ice could be photodesorbed by stellar radiation. Dust evolution models predict that these large icy grains will drift radially towards the star faster than the PPD gas, enhancing the gaseous carbon and oxygen abundances inside the CO and water snowlines \citep{du15, krijt18}, as seen for HD 163296 (Zhang et al. 2020). In contrast, TW Hya does not show a strong gaseous volatile molecule enhancement inside its CO snowline at 20 AU \citep{zhang2017}. Either CO ice is chemically processed into a more rocky carbon carrier or the radial drift of solids is inefficient due to grains collecting in traps \citep{bosman2019}, which may be the first step to forming planets, or planetesimals themselves. 

Measurements of the bulk gas elemental abundances interior to the dust sublimation rim can probe whether the icy C and O have returned to the gas phase or are still missing. The former would imply chemical processing of the ice into other ice species with higher sublimation temperatures, while the latter would suggest that the C and O have been locked into midplane ice reservoirs. For stars >1.4 M$_\odot$, the inner disk elemental abundances can be inferred from the stellar photospheric abundances, due to the formation of a radiative outer envelope that isolates recently accreted material at the stellar surface \citep{kama2015, jermyn2018}. However, this technique does not work with PPDs around lower mass T Tauri stars, which are the precursors of solar type stars, as their outer layers are convective. Interior circulation thus mixes away the compositional signatures of their accreted gas. Recently, we directly measured the inner PPD gas around a set of these lower mass stars through the use of near-infrared atomic C emission lines \citep{mcclure2019}. That work inferred carbon locking in these PPDs but did not observationally measure abundances of hydrogen or other critical elements, e.g. O or Si, which could identify whether these elements were locked in rocky versus icy materials.

In this Letter, we model the hydrogen density and atomic abundances for the inner disk of TW Hya from near-infrared atomic emission lines. We then use these abundances to identify the locations where rock- and ice-forming elements are being locked into midplane dust grain reservoirs in this system. 

%Here we subtract multiple epochs of near-infrared (NIR) emission line spectra of TW Hya to isolate a clump of hot gas inside the dust sublimation radius. By fitting emission line flux ratios in the data using the photoionization and chemistry code Cloudy \citep{ferland2017}, we derive the density of hydrogen directly and determine absolute abundances of several rock- and ice-forming elements. We then reconstruct the bulk elemental abundances of solids left behind in TW Hya's disk and compare both of these results with elemental abundances of solar system bodies to determine TW Hya's rocky solids fraction and locking efficiency.

%-------------------------------------- Two column figure (place early!)
   \begin{figure*}
   \centering
   \includegraphics[width=8.5cm, angle=90]{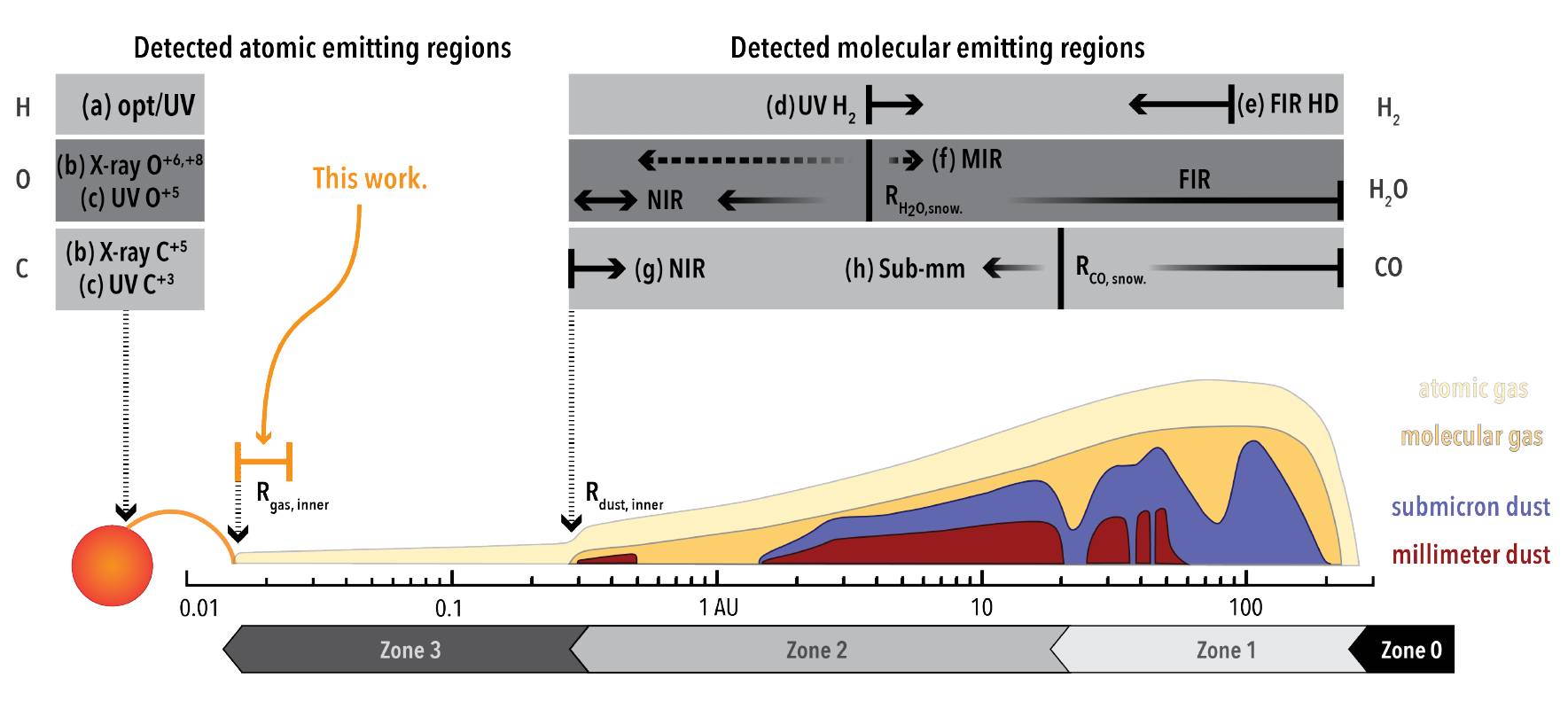}
   \caption{{\bf Schematic of emitting regions for previous TW Hya observations, compared with the region probed in this work.} For comparison, observed disk structures in millimeter grains (ALMA,     \citep{andrews16}, VLTI, \citep{menu14}), micron-sized grains (SPHERE, \citep{vanboekel2017}), and CO gas \citep{banzatti2015,huang18} are shown. The vertical scale is not meaningful. {\bf References:} (a) \citet{dupree12}, (b) \citet{brickhouse10}, (c) \citet{herczeg02}, (d) \citet{france2012b}, (e) \citet{bergin+13}, (f) \citet{zhang13}, \citet{bosman2019}, (g) \citet{banzatti2015}, \citet{bosman2019}, (h) \citet{zhang2017}. The `zone' labels correspond to the regions described in the analytic calculation of \S\ref{outercn}.}
    \label{fig1}%
    \end{figure*}
%--------------------------------------------------------------------

%--------------------------------------------------------------------------------------------------------------------------------------------------------------------------------------------------------------------------------------------------------
%--------------------------------------------------------------------------------------------------------------------------------------------------------------------------------------------------------------------------------------------------------
%--------------------------------------------------------------------------------------------------------------------------------------------------------------------------------------------------------------------------------------------------------
\section{Data analysis and physical scenario}
\label{analysis}
To measure the \textbf{disk} abundances inside the dust sublimation radius, we analyze spectra of TW Hya spanning 0.8 - 2.5 $\mu$m taken in three epochs: 2010a and 2010b (VLT X-shooter), and 2013 (Magellan FIRE). The observations and data reduction details are given in Appendix \ref{obsdat}; briefly, the reduced spectra were corrected for telluric absorption, heliocentric velocities, and convolved down to a uniform spectral resolution, R$\sim$6000. All three epochs show H$^0$, He$^0$, Ca$^{+1}$, and O$^0$ emission lines (Figs. \ref{fluxes_1} and \ref{fluxes_2}). In the 2013 spectrum, these lines were brighter and showed additional emission from C$^0$ and S$^0$ lines at 1.07 and 1.045 $\mu$m, respectively, as shown in Figures \ref{fluxes_1} and \ref{fluxes_2}. We subtract the continuum under the lines of interest in each spectrum using the automated continuum fitter from \citet{mcclure+13a}. For the detected Ca$^{+1}$ , O$^0$, C$^0$, and S$^0$ lines, we integrated the lines and list the fluxes in Table \ref{tab_flux_abs}. For the non-detected strong lines, we use the continuum rms value to obtain 3$\sigma$ upper limits. 

\begin{figure*}
   \centering
   \includegraphics[width=15cm]{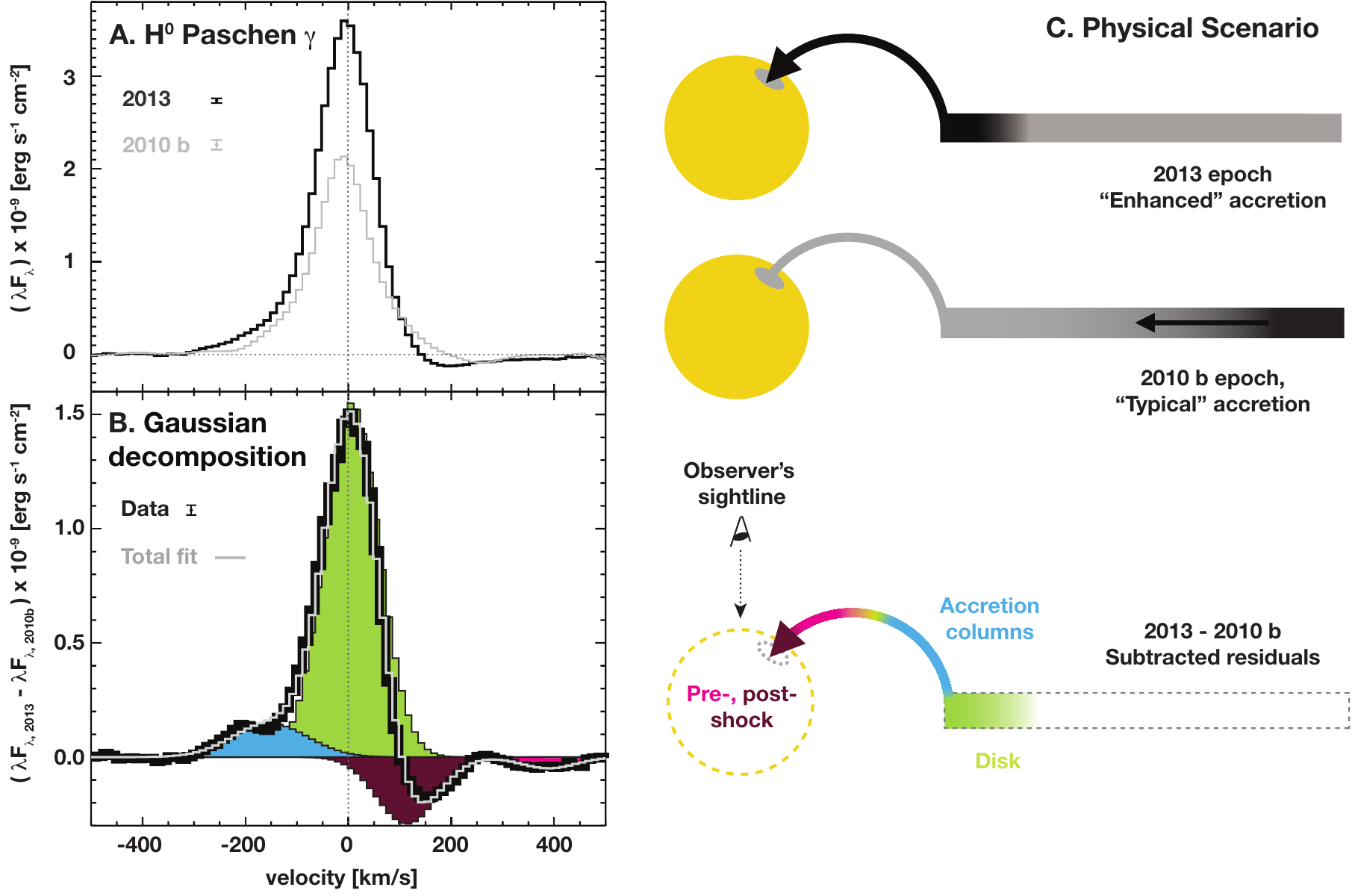}
   \caption{{\bf Isolation of the hydrogen contribution from the inner disk edge.} {\bf A. Comparison of the Pa$\gamma$ line emission} for the 2010b epoch (gray) and 2013 epoch (black) after continuum subtraction. Dashed vertical line indicates the local stellar velocity, to which the spectra have been corrected. {\bf B. Gaussian decomposition fit} (gray) of the residual spectrum (black) obtained by subtracting the 2010b epoch spectrum from that of 2013. Contributions are indicated with colored fill and include blue-shifted emission (blue: -152 km/s, FWHM$\sim$172 km/s ), a centrally peaked emission (green: 7 km/s, FWHM$\sim$126 km/s), and two redshifted absorption components (dark red: 114 km/s, FWHM$\sim$137 km/s; bright red: 389 km/s, FWHM$\sim$124 km/s). {\bf C. Cartoon of the physical scenario} revealed by the variability and gaussian fitting.}
  \label{pagamma_fit}%
  \end{figure*}

The strengthening of all emission lines in 2013 suggests an increase in the density of the accreting, inner disk  material for this epoch. We use the H recombination line profiles to confirm this scenario. As shown in Fig. \ref{pagamma_fit}, the continuum-subtracted Paschen $\gamma$ lines from 2010b and 2013 are both centrally peaked at the stellar rest velocity, with inverse P-Cygni profiles in the wings. However, the maxima and minima of the 2013 profile are stronger. By subtracting the 2010 line profiles from the 2013 spectrum, we remove the contributions of the star and bulk disk to the emission lines. We model the residual profile with a four-component Gaussian decomposition.  A centrally peaked component is found, within the observed velocity resolution uncertainty. Since any stellar chromospheric contribution was removed during the subtraction of the weaker accretion epoch, our fitting results suggests the centrally peaked component could only come from the inner accretion disk, for which the Keplerian motion is in the plane of the sky. In contrast, there is additional inverse P-Cygni emission peaking at -152 km/s and absorbing at +114 km/s. Since TW Hya is face-on \citep{andrews16}, we interpret the blue-shifted emission component as the contribution from magnetospheric accretion columns launching out of the disk plane and the red-shifted absorption component to the post-accretion shock braking region at the stellar surface. A fourth Gaussian component absorbs at +389 km/s, close to the predicted free-fall velocity at the accretion shock interface and $\sim$4 times the post-shock velocity, as expected theoretically \citep{hartmann16}. Therefore the shape of the H line profiles support our hypothesis of an increase in the density of the inner disk gas. We model the physical conditions and elemental abundances in this gas in the section below.

%------------------- FIGURE ------------------- 
%---------------------------------------------------------  

%--------------------------------------------------------------------------------------------------------------------------------------------------------------------------------------------------------------------------------------------------------
%--------------------------------------------------------------------------------------------------------------------------------------------------------------------------------------------------------------------------------------------------------
%--------------------------------------------------------------------------------------------------------------------------------------------------------------------------------------------------------------------------------------------------------

\section{Modeling}
The gaussian decomposition fit to the H line profiles kinematically isolates the 2013 inner disk hydrogen emission. Using an existing 1D model of the inner disks of T Tauri stars \citep{mcclure2019} driven by the photoionization and chemistry code Cloudy \citep{ferland2017}, we compare the observed flux ratio between the H Pa$\beta$ and Pa$\gamma$ emission lines to measure observationally the H number density, $n_{\rm H}$, in the inner disk. Then we model the observed flux ratios between these elemental lines and Pa$\gamma$, together with the derived $n_{\rm H}$, to measure total elemental abundances relative to the total amount of H for the inner disk gas.

Our 1D model assumes the inner disk extends from the co-rotational radius at 0.015 AU out to 0.024 AU. This width was determined from the radial distance travelled on a viscous timescale of $\sim$2.5 years, the difference between epochs 2013 and 2010b, assuming a `typical' $\alpha$=0.01. We took a radially dependent power-law hydrogen density of power -2. We created a composite radiation field to represent the joint contributions from the star and accretion shock. For the star we used ATLAS9 stellar models with solar metallicity \citep{castelli03} at the stellar effective temperature and luminosity, T$_{eff}$ = 3850 K with L$_{*}$ = 0.3 L$_{\odot}$. We treated the accretion shock as a combination of UV and x-ray emission. The UV emission is approximated by a stellar emission spectrum with a temperature and luminosity, T$_{\rm shock}$ = 8000 K with L$_{\rm shock}$ = 0.01 L$_{\odot}$. We include a hard X-ray component, assuming a 10 MK bremsstrahlung emission spectrum, with a luminosity of L$_X$=10$^{30}$ erg/s between 0.3 keV and 10 keV \citep{rab18}. Accretion heating was enabled using the Hextra keyword, as described in Section 11.9.4 of the Cloudy user manual, assuming $\alpha$=0.01, the stellar mass, M$_{*}$=0.6 M$_{\rm \odot}$ \citep{sokal2018}, and the radius set to 0.024 AU.

Cloudy includes a chemical network with atomic and molecular opacities appropriate to the hot, dense, dust-free region interior to the inner dust rim, which enables radiative transfer calculation in the absence of dust opacity. It is important to note that due to the high densities at this location, the chemistry does not behave as in a typical PDR, i.e. ionization of H is provided by charge exchange collisions, rather than photo- or collisional-ionization. Therefore C$^0$ and H$^0$ coexist and are the main C and H carriers despite the exposure to UV and x-rays or expected self-shielding by H$_2$. The inner disk chemistry is discussed extensively in the text of Section 3 and Figures 3 and 4 of \citet{mcclure2019}, but we also summarize briefly the salient mechanisms, sources of systematic uncertainty in the final abundances, and a complete list of the model parameters in Appendix B.
 
A value of log($n_{\rm H}$) = 14.40 cm$^{-3}$ at the inner edge of the disk model is required to fit the observed Pa$\beta$/Pa$\gamma$ = 2.3 flux ratio in the inner disk kinematic component from 2010b-2010a. By 2013, the inner disk gas density increased to log (n$_{\rm H}$) = 14.74 cm$^{-3}$. This is consistent with the physical scenario derived from the hydrogen line profiles. We then fixed log$\left(n_{\rm H}\right)$ to the 2013 value of 14.74 cm$^{-3}$ and varied the atomic abundances of C, O, S, and Ca. Plots of the model fluxes with elemental abundance for each element are shown in Fig. \ref{fluxes_3}. After confirming that the Pa$\beta$/P$\gamma$ ratio was still valid, we determined 3$\sigma$ upper limits to the abundances of N, K, Na, Mg, Fe, Al, and Si, which have undetected strong lines in this region. The resulting abundances are given in Table \ref{tab_flux_abs}.

%--------------------------------------------------------------------------------------------------------------------------------------------------------------------------------------------------------------------------------------------------------
%--------------------------------------------------------------------------------------------------------------------------------------------------------------------------------------------------------------------------------------------------------
%--------------------------------------------------------------------------------------------------------------------------------------------------------------------------------------------------------------------------------------------------------

\section{Results and discussion}

The results of this comprehensive elemental abundance measurement for the inner disk of a T Tauri star contradict the prediction of dust evolution models. {\it All measured elements are depleted with respect to the sun, with volatiles depleted by up to two orders of magnitude and refractories by up to five orders of magnitude.} We note that our C/H and O/H values of 3.39$\times$10$^{-6}$ and 1.60$\times$10$^{-5}$, respectively, are consistent within our systematic uncertainties with the C/H and O/H values for TW Hya recently found by \citet{bosman2019}: 3$\times$10$^{-6}$ and 6$\times$10$^{-6}$, respectively. Likewise, our N/H upper limit of $<$1.6$\times$10$^{-5}$ is depleted by a factor of $\sim$4. This value encompasses the depleted N/H of $\sim$3$\times$10$^{-6}$ inferred for TW Hya by \citet{vanthoff2017} from ALMA observations of N$_2$H$^+$. It is also consistent with an earlier finding of N depletion on the stellar surface of TW Hya, ranging between factors of 1.5 and 5 \citep{brickhouse10}. As shown in Fig. \ref{abunds}, the accretion of primary atmospheres from this gas would not explain the enhanced giant planet atmospheric volatile abundances seen in our solar system. However, the abundance pattern seen between the elements in TW Hya's gas is a close inverse of the CI chondrite abundance pattern, suggesting that the depletion could be caused by the retention of chondrite-like solids in the disk.

\subsection{Outer disk dust traps for C, N, and S}
\label{outercn}
\citet{bosman2019} suggest that the C depletion in TW Hya's inner disk could result from either a dust trap at the CO snowline ($\sim$20 AU) or a CO$_2$ trap close to the H$_2$O snowline (4 AU). Due to the fact that our C/H measurement is entirely within the dust sublimation rim, we can use previous measurements of CO and a simple analytic treatment (Appendix \ref{c_lock}) to estimate the location where C is locked out. 

First, we divide the disk into three zones (see Fig. \ref{fig1}), one exterior to the CO snowline (Zone 1), a second between the CO snowline and dust inner rim (Zone 2), and a third interior to the dust inner rim (Zone 3). This division closely matches the dust traps found by \citet{bosman2019}, with the dust traps at the end of Zone 1 and in the middle of Zone 2. For each zone we take a representative value of C/H. In Zone 3, a C/H depletion factor of 79 is taken from the present work. For Zones 1 and 2, we calculate a C/H ratio from the C$^{18}$O measurements of \citet{zhang2019}. That works finds a minimum depletion factor for CO/H$_2$ of $\sim$78 outside of the CO snowline, while the average depletion factor within the CO snowline is $\sim$20, taken relative to the interstellar medium CO/H$_2$ value of $10^{-4}$. These values convert into C/H values of 6.4$\times10^{-7}$ and 2.5$\times10^{-6}$, making the gross approximation that all gas phase C is in CO.

Then we consider that an infinitely thin parcel of gas and dust is transported from the outer edge of Zone 1 through each zone sequentially. The dust is composed of a CO ice component and a "higher temperature" component with carbon carriers that sublimate at a higher temperatures than CO, e.g. CO$_2$, organic residues, or graphite. The dust is either locked into the disk in a given zone or allowed to pass through into the next zone, where a portion of it sublimates. We assume a higher temperature component fraction of 0.83, which is the refractory C fraction observed in comet 67P \citep{rubin19}, and use Eqs. \ref{fl1} and \ref{fl2} to calculate the fraction of the initially available carbon dust that is locked into the disk the parcel moves through each zone. We find that 95.9\% of the initial carbon dust mass is locked into the disk exterior to the CO snowline (Zone 1). Therefore the combined efficiency of any dust traps exterior to that point is also $\sim$96\%. Only 4.1\% of the initial dust is transported into Zone 2, with 0.7\% of that dust in volatiles that sublimate. Between the CO snowline and the inner dust sublimation rim (Zone 2), an additional 3.06\% of the initial dust mass is locked. Only 0.34\% of the initial dust crosses the inner dust rim into Zone 3, where it sublimates. In total, only 1.04\% of the initial Zone 1 carbon dust mass is returned to the gas phase to be accreted onto the central star. The agreement between our C/H abundance in Zone 3 and that of \citet{bosman2019}, which is representative of the gap at 1 AU in Zone 2, between the H$_2$O and CO$_2$ ice snowlines and the silicate dust trap, suggests that the 3\% of initial dust locked in Zone 2 is located beyond the 1 AU gap. The efficiency of locking in traps between 1.5 and 20 AU is then 90\%. 

Most of the uncertainty in the above estimate depends on the higher temperature carbon fraction. Reasonable values range from 0.5, which would lock 99\% of the initial carbon mass within Zone 1, at the expense of Zone 2. Decreasing the CO/H$_2$ depletion factor from 20 to 15, the value observed by \citet{zhang2019} at 4 AU, would reduce the locked dust mass beyond the CO snowline to 94\% of the initial dust mass, suggesting that even if CO is converted into CO$_2$ ice, most of that material is already trapped efficiently in the outer disk rings. Locking of solids chemically isolates the inner disk, preventing replenishment of higher temperature C and potentially leading to Earth-like C abundances \citep{klarmann18, anderson17} for forming terrestrial planets.

A quantative analysis of the N depletion validates our C transport model and choice of $f_r$. The N/H found by \citet{vanthoff2017} is only $\sim$4\% of the solar abundance, i.e. \emph{the same fraction of N is missing as we find for C beyond the CO snowline in our model using $f_r$=0.83.} Since N$_2$ is likely the dominant solid N-carrier in the outer disk, and its midplane snowline at $\sim$18 K \citep{piso2016} should be just outside of the CO snowline, then the N$_2$ must also be trapped in the three submillimeter rings beyond 30 AU.

In contrast with TW Hya, several other disks show no evidence for C or N gas depletion. HD 163296 has enhanced C/H gas interior to the CO snowline \citep{zhang2020} due to radial drift, while IM Lup shows depletions of \emph{at most} a factor of 4 \citep{cleeves2018}. For IM Lup, the gas temperature structure puts the N$_2$ snowline at >250 AU \citep[Fig. 9]{cleeves2016c}, which is well outside of the outermost dust ring in IM Lup at 134 AU \citep{huang18}. Therefore little N$_2$ ice is trapped in this disk, and gaseous N is not as depleted. A similar lack of dust traps beyond the N$_2$ snowline could account for the ISM N abundance found by \citet{anderson2019} for a set of $\sim$5-11 Myr disks in upper Sco; however these disks lack the same high spatial resolution continuum observations that enabled the ring detections in IM Lup and TW Hya. For HD 163296, a ring with a deep gap does sit at 100 AU, just outside the CO snowline at 70 AU, that apparently does not trap dust efficiently \citep{isella2018}. However, while TW Hya's outer disk gaps are less deep than in HD 163296, it does have three, which may explain their large cumulative trapping efficiency. Alternatively, if the dust in the rings of TW Hya has grown larger than in HD 163296, it may be filtered out more efficiently \citep{rice2006}, or planetesimals have formed, then they would no longer drift efficiently regardless of trapping.

\begin{figure}
  % \centering
   \includegraphics[width=9cm]{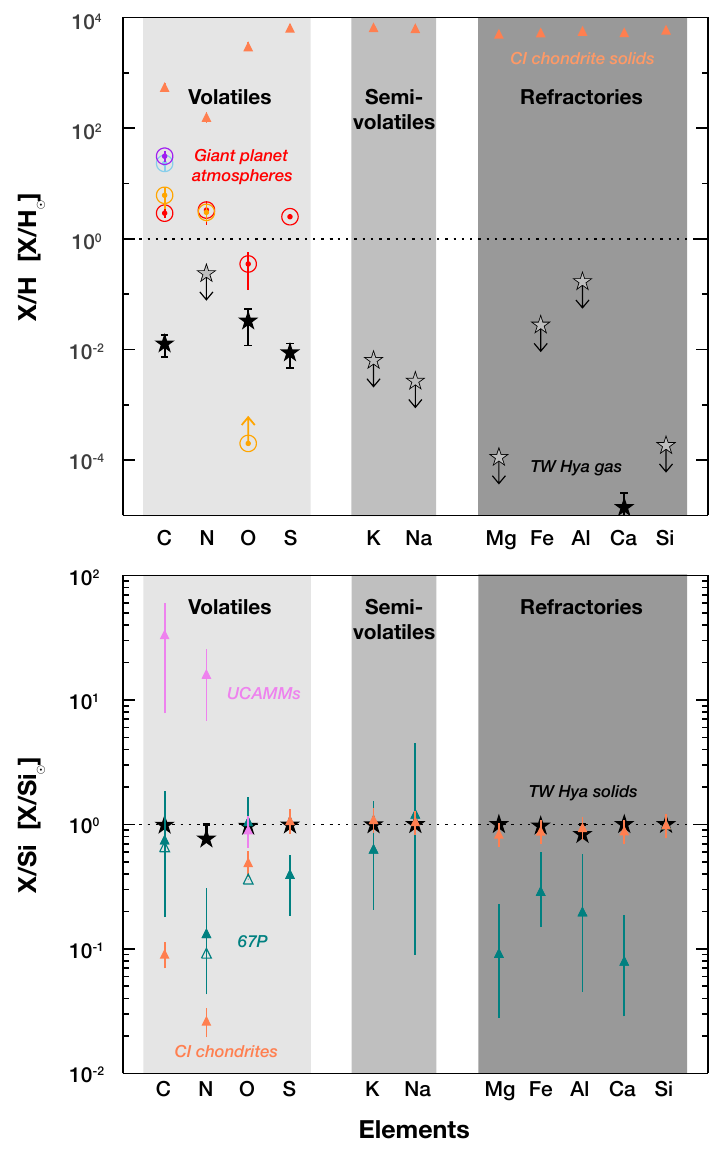}
   \caption{{\bf Top} Comparison of bulk elemental gas abundances inside TW Hya's dust sublimation rim (black stars, gray stars with arrows for upper limits), with atmospheric abundances of \citep{owen03}, Jupiter (red circles), Saturn (orange circles), Uranus (light blue circle), and Neptune (purple circle), and CI chondrite solids \citep[][salmon triangles]{lodders2010}. All data are plotted with error bars. Abundances are with respect to hydrogen and are normalized to the solar abundances \citep{asplund2009}, with the solar value indicated by the dotted horizontal line. {\bf Bottom} Comparison of bulk elemental abundances of solids for: TW Hya (black stars), CI chondrites \citep[][salmon triangles]{lodders2010}, comet 67P (bulk, dust/ice ~ 1, teal triangles) \citep{bardyn17,rubin19}, and UCAMMs (violet triangles) \citep{dartois2017,mathurin19}}
  \label{abunds}%
  \end{figure}

As a further check of the locking location, in the bottom panel of Fig. \ref{abunds} we compare the bulk atomic composition of the solids left behind in TW Hya's disk with the abundances of some primitive solar system bodies originating roughly in each of the three zones, e.g. CI chondrites \citep[Zone 3]{lodders2010}, comet 67P \citep[Zone 2] {bardyn17,rubin19}, and ultra-carbonaceous Antarctic micro-meteorites \citep[][UCAMMs, Zone 1]{dartois2017,mathurin19}. The TW Hya solids are all nearly solar, and within the uncertainties, chondritic abundances are generally a good match to the refractory and semi-volatile elements, as well as S. However, chondrites underproduce TW Hya's C, N, and O abundances, by an order of magnitude in the case of the former two. Comet 67P and the UCAMMs are more abundant in all three of these elements, but with opposite abundance patterns, i.e. low N/O in comets and high N/O in UCAMMs.  A crude check fitting a blend of abundances from CI chondrites and either Comet 67P or UCAMMs, or both combined, to TW Hya's solid abundances confirms that neither 67P nor UCAMMs on their own provide a good fit to C, N, and O. The former overproduces O and underproduces N, while the latter underproduces N and O, and both fits are poor overall ($\chi^2\sim$5000-8000). Interestingly, a combination of all three bodies succeeds in fitting C and O, with a $\chi^2\sim$3, but not N. 

There are several potential explanations of the apparent `excess' nitrogen in TW Hya's solids. The UCAMMs are thought to originate from differentiated parent bodies beyond the N$_2$ snowline, in which N$_2$ was converted to higher temperature residues \citep{dartois2017}, but they are ultimately recovered in Antarctica. These residues are more stable against sublimation than the original N$_2$ ice, but is is possible that they encountered some loss on atmospheric entry, artificially lowering the observed abundances. Alternatively, there may be differences between the dust trapping patterns in the solar system and TW Hya that produce different types of primitive bodies; i.e. CI chondrites, comet 67P, and UCAMMs may not be an appropriate basis set for TW Hya's solids. This is generally true to some extent, particularly as we have only used one comet, 67P, to represent the entire cometary class, due to a lack of comprehensive elemental abundances for additional comets. Finally, the primitive bodies represent the cumulative elements locked into solids at a particular disk radius over the lifetime of the solar nebula, while TW Hya's solid abundances represent a snapshot of the material that has been locked out of a moving parcel of gas over multiple radii and a later range of times, from $\sim$2-8 Myr based on viscous accretion timescales. If the N-rich dust trap formed later in TW Hya's lifetime than it did in the solar nebula, then the net effect on the cumulative N abundances in the primitive bodies would be diluted.

Unlike N, S can be found in both volatile molecules, e.g. H$_2$S, or minerals that contain more refractory elements, i.e. FeS. Analysis of the atmospheres of accreting Herbig AeBe stars with disks suggests that of the total S content, $<$1\% is in gaseous form, with 89$\pm$8\% of S is in refractory minerals \citep{kama2019} and the remainder in ices. The depletion of S here is consistent with an 89\% refractory fraction; if the trap at 2 AU lets through 0.3\% of the total initial dust mass, as for the C transport model, then sublimation of volatile ice with an initial fraction of 11\% in that dust can account for all of the gas phase S that we see in the inner disk. The mineral sulphides pass through the 1 AU gap into the inner, refractory disk, which is discussed below.

\subsection{Inner disk refractory dust trap}

One of the refractory elements, Ca, was firmly detected in the gas phase with a depletion of nearly five orders of magnitude, relative to the solar value.  While more depletion of refractories than volatiles is to be expected, due to rocky cores in all icy dust grains, at first glance five orders of magnitude seems extreme. However, two of the upper limits to the abundances of the other refractory elements, Mg and Si, are depleted by approximately 10$^{-4}$, roughly supporting such low values. The latter is qualitatively consistent with previous findings of weaker Si relative to C or N in TW Hya's accretion streams (Ardila et al. 2013). Our extreme depletions suggest that there is an additional dust trap interior to the sublimation point of most species considered for our higher temperature carbon component, where the bare silicate grains are prevented from sublimating. 

An obvious choice for this trap is the dust ring interior to 0.5 AU, detected with ALMA \citep{andrews16}. Inner disk temperature and density structures calculated self-consistently in hydrostatic equilibrium normally reach the silicate sublimation temperature of 1400 K at $\sim$0.1 AU, in M0 T Tauri stars with mass accretion rates $>$10$^{-9}$ M$_{\odot}$yr$^{-1}$ \citep[e.g. GO Tau in][Fig. 10]{mcclure+13b}. However, using near- and mid-IR interferometry \citet{menu14} find the inner rim for TW Hya at 0.3 AU instead, suggesting that the inner edge is associated with a pressure maximum rather than the silicate sublimation front. Blocking accretion or drift of large silicate grains at 0.3 AU would retain the elements Mg, Si, and Ca in the disk. The cumulative amount of the initial dust mass passing through the three submillimeter rings outside of the CO and N$_2$ snowlines is $\sim$4\%, and the trap near the water snowline only allows 10\% of that first 4\% of the dust to continue through. To reproduce the fraction of initial Si, Mg, and Ca returned to the inner disk gas of $\sim$10$^{-4}$-2$\times$10$^{-5}$, only 4.6-0.6\% of the dust in the inner 0.3-0.5 AU can pass through the 0.3 AU dust trap and sublimate. This suggests that the trapping efficiency is inversely proportion to the disk radius, with 96\% trapping efficiency cumulatively over the outer three submillimeter rings, 90\% at the 2 AU ring, and $\sim$95-99\% at the 0.3 AU ring.

If the refractory C within the 0.3-0.5 AU ring is efficiently removed by photolysis or oxidation, as suggested by \citet{anderson17}, then the high trapping efficiency of the 2 AU and outer submillimeter dust traps cuts off C replenishment by drifting solids, as proposed by \citet{klarmann18}. Should the innermost, refractory trap of TW Hya eventually form terrestrial planets, they would form from very dry and poor in C and N dust, possibly resulting in Earth-like depletions of these elements.

%------------------- FIGURE ------------------- 
%---------------------------------------------------------  

\section{Conclusions}
To summarize, we piloted a novel method to determine the bulk atomic composition of unseen solids locked at the midplanes of protoplanetary disks, using the TW Hya disk. The volatile elements in this system are depleted by a factor of $\sim$100 with respect to the solar value, while the refractories are depleted by up to five orders of magnitude. The high degree of depletion and difference between the depletion factors of the volatile and refractory elements can be explained if the submillimeter rings all host dust traps, including the truncated ring from 0.3-0.5 AU.

A simple model for the efficiency of dust retention in these traps finds that the inner two dust traps at 0.3 and 2 AU have a high efficiency ($\ge$90\% each), while the three submillimeter rings beyond the midplane CO and N$_2$ snowlines are collectively $\sim$96\%, less individually. Such a high retention efficiency, relative to other disks, may be due to the positioning of the rings relative to individual molecular snowlines as well as the age and maximum grain size in the system. In any event, only a small fraction of the elements in dust end up accreted onto the central star. These results, combined with theoretical modeling of refractory C destruction, suggest that any terrestrial planets forming in the trap within TW Hya's inner disk from the remaining solids will be relatively `dry' and C-poor, similar to those in our solar system.

\begin{acknowledgements}
This publication is part of a project that has received funding from the European Union's Horizon 2020 research and innovation program under the Marie Sk$\l$odowska-Curie grant agreement ICED No. 749864 (M.M.). M.K. was supported by the University of Tartu ASTRA project 2014-2020.4.01.16-0029 KOMEET, financed by the EU European Regional Development Fund. We thank the anonymous referee for comments that significantly improved the manuscript.
\end{acknowledgements}

 %  \bibliography{References} % your references Yourfile.bib

\begin{thebibliography}{57}
%\expandafter\ifx\csname natexlab\endcsname\relax\def\natexlab#1{#1}\fi

\bibitem[{{Anderson} {et~al.}(2017){Anderson}, {Bergin}, {Blake}, {Ciesla},
  {Visser}, \& {Lee}}]{anderson17}
{Anderson}, D.~E., {Bergin}, E.~A., {Blake}, G.~A., {et~al.} 2017, \apj, 845,
  13

\bibitem[{{Anderson} {et~al.}(2019){Anderson}, {Blake}, {Bergin}, {Zhang},
  {Carpenter}, {Schwarz}, {Huang}, \& {{\"O}berg}}]{anderson2019}
{Anderson}, D.~E., {Blake}, G.~A., {Bergin}, E.~A., {et~al.} 2019, \apj, 881,
  127

\bibitem[{{Andrews} {et~al.}(2016){Andrews}, {Wilner}, {Zhu}, {Birnstiel},
  {Carpenter}, {P{\'e}rez}, {Bai}, {{\"O}berg}, {Hughes}, {Isella}, \&
  {Ricci}}]{andrews16}
{Andrews}, S.~M., {Wilner}, D.~J., {Zhu}, Z., {et~al.} 2016, \apjl, 820, L40

\bibitem[{{Asplund} {et~al.}(2009){Asplund}, {Grevesse}, {Sauval}, \&
  {Scott}}]{asplund2009}
{Asplund}, M., {Grevesse}, N., {Sauval}, A.~J., \& {Scott}, P. 2009, \araa, 47,
  481

\bibitem[{{Banzatti} \& {Pontoppidan}(2015)}]{banzatti2015}
{Banzatti}, A. \& {Pontoppidan}, K.~M. 2015, \apj, 809, 167

\bibitem[{{Bardyn} {et~al.}(2017){Bardyn}, {Baklouti}, {Cottin}, {Fray},
  {Briois}, {Paquette}, {Stenzel}, {Engrand}, {Fischer}, {Hornung}, {Isnard},
  {Langevin}, {Lehto}, {Le Roy}, {Ligier}, {Merouane}, {Modica},
  {Orthous-Daunay}, {Ryn{\"o}}, {Schulz}, {Sil{\'e}n}, {Thirkell}, {Varmuza},
  {Zaprudin}, {Kissel}, \& {Hilchenbach}}]{bardyn17}
{Bardyn}, A., {Baklouti}, D., {Cottin}, H., {et~al.} 2017, \mnras, 469, S712

\bibitem[{{Bergin} {et~al.}(2015){Bergin}, {Blake}, {Ciesla}, {Hirschmann}, \&
  {Li}}]{bergin15}
{Bergin}, E.~A., {Blake}, G.~A., {Ciesla}, F., {Hirschmann}, M.~M., \& {Li}, J.
  2015, Proceedings of the National Academy of Science, 112, 8965

\bibitem[{{Bergin} {et~al.}(2013){Bergin}, {Cleeves}, {Gorti}, {Zhang},
  {Blake}, {Green}, {Andrews}, {Evans}, {Henning}, {{\"O}berg}, {Pontoppidan},
  {Qi}, {Salyk}, \& {van Dishoeck}}]{bergin+13}
{Bergin}, E.~A., {Cleeves}, L.~I., {Gorti}, U., {et~al.} 2013, \nat, 493, 644

\bibitem[{{Bosman} \& {Banzatti}(2019)}]{bosman2019}
{Bosman}, A.~D. \& {Banzatti}, A. 2019, \aap, 632, L10

\bibitem[{{Brickhouse} {et~al.}(2010){Brickhouse}, {Cranmer}, {Dupree}, {Luna},
  \& {Wolk}}]{brickhouse10}
{Brickhouse}, N.~S., {Cranmer}, S.~R., {Dupree}, A.~K., {Luna}, G.~J.~M., \&
  {Wolk}, S. 2010, \apj, 710, 1835

\bibitem[{{Castelli} \& {Kurucz}(2003)}]{castelli03}
{Castelli}, F. \& {Kurucz}, R.~L. 2003, in IAU Symposium, Vol. 210, Modelling
  of Stellar Atmospheres, ed. N.~{Piskunov}, W.~W. {Weiss}, \& D.~F. {Gray},
  A20

\bibitem[{{Cleeves} {et~al.}(2016){Cleeves}, {{\"O}berg}, {Wilner}, {Huang},
  {Loomis}, {Andrews}, \& {Czekala}}]{cleeves2016c}
{Cleeves}, L.~I., {{\"O}berg}, K.~I., {Wilner}, D.~J., {et~al.} 2016, \apj,
  832, 110

\bibitem[{{Cleeves} {et~al.}(2018){Cleeves}, {{\"O}berg}, {Wilner}, {Huang},
  {Loomis}, {Andrews}, \& {Guzman}}]{cleeves2018}
{Cleeves}, L.~I., {{\"O}berg}, K.~I., {Wilner}, D.~J., {et~al.} 2018, \apj,
  865, 155

\bibitem[{{Dartois} {et~al.}(2017){Dartois}, {Chabot}, {Pino}, {B{\'e}roff},
  {Godard}, {Severin}, {Bender}, \& {Trautmann}}]{dartois2017}
{Dartois}, E., {Chabot}, M., {Pino}, T., {et~al.} 2017, \aap, 599, A130

\bibitem[{{Drake} {et~al.}(2005){Drake}, {Testa}, \& {Hartmann}}]{drake2005}
{Drake}, J.~J., {Testa}, P., \& {Hartmann}, L. 2005, \apj, 627, L149

\bibitem[{{Du} {et~al.}(2015){Du}, {Bergin}, \& {Hogerheijde}}]{du15}
{Du}, F., {Bergin}, E.~A., \& {Hogerheijde}, M.~R. 2015, \apjl, 807, L32

\bibitem[{{Dupree} {et~al.}(2012){Dupree}, {Brickhouse}, {Cranmer}, {Luna},
  {Schneider}, {Bessell}, {Bonanos}, {Crause}, {Lawson}, {Mallik}, \&
  {Schuler}}]{dupree12}
{Dupree}, A.~K., {Brickhouse}, N.~S., {Cranmer}, S.~R., {et~al.} 2012, \apj,
  750, 73

\bibitem[{{Ferland} {et~al.}(2017){Ferland}, {Chatzikos}, {Guzm{\'a}n},
  {Lykins}, {van Hoof}, {Williams}, {Abel}, {Badnell}, {Keenan}, {Porter}, \&
  {Stancil}}]{ferland2017}
{Ferland}, G.~J., {Chatzikos}, M., {Guzm{\'a}n}, F., {et~al.} 2017, \rmxaa, 53,
  385

\bibitem[{{France} {et~al.}(2012){France}, {Schindhelm}, {Herczeg}, {Brown},
  {Abgrall}, {Alexander}, {Bergin}, {Brown}, {Linsky}, {Roueff}, \&
  {Yang}}]{france2012b}
{France}, K., {Schindhelm}, E., {Herczeg}, G.~J., {et~al.} 2012, \apj, 756, 171

\bibitem[{{Hartmann} {et~al.}(2016){Hartmann}, {Herczeg}, \&
  {Calvet}}]{hartmann16}
{Hartmann}, L., {Herczeg}, G., \& {Calvet}, N. 2016, \araa, 54, 135

\bibitem[{{Heays} {et~al.}(2014){Heays}, {Visser}, {Gredel}, {Ubachs}, {Lewis},
  {Gibson}, \& {van Dishoeck}}]{heays2014}
{Heays}, A.~N., {Visser}, R., {Gredel}, R., {et~al.} 2014, \aap, 562, A61

\bibitem[{{Herczeg} {et~al.}(2002){Herczeg}, {Linsky}, {Valenti},
  {Johns-Krull}, \& {Wood}}]{herczeg02}
{Herczeg}, G.~J., {Linsky}, J.~L., {Valenti}, J.~A., {Johns-Krull}, C.~M., \&
  {Wood}, B.~E. 2002, \apj, 572, 310

\bibitem[{{Hily-Blant} {et~al.}(2019){Hily-Blant}, {Magalhaes de Souza},
  {Kastner}, \& {Forveille}}]{hily-blant2019}
{Hily-Blant}, P., {Magalhaes de Souza}, V., {Kastner}, J., \& {Forveille}, T.
  2019, \aap, 632, L12

\bibitem[{{Hogerheijde} {et~al.}(2011){Hogerheijde}, {Bergin}, {Brinch},
  {Cleeves}, {Fogel}, {Blake}, {Dominik}, {Lis}, {Melnick}, {Neufeld},
  {Pani{\'c}}, {Pearson}, {Kristensen}, {Y{\i}ld{\i}z}, \& {van
  Dishoeck}}]{hogerheijde+11}
{Hogerheijde}, M.~R., {Bergin}, E.~A., {Brinch}, C., {et~al.} 2011, Science,
  334, 338

\bibitem[{{Huang} {et~al.}(2018){Huang}, {Andrews}, {Cleeves}, {{\"O}berg},
  {Wilner}, {Bai}, {Birnstiel}, {Carpenter}, {Hughes}, {Isella}, {P{\'e}rez},
  {Ricci}, \& {Zhu}}]{huang18}
{Huang}, J., {Andrews}, S.~M., {Cleeves}, L.~I., {et~al.} 2018, \apj, 852, 122

\bibitem[{Isella {et~al.}(2018)Isella, Huang, Andrews, Dullemond, Birnstiel,
  Zhang, Zhu, Guzmán, Pérez, Bai, \& et~al.}]{isella2018}
Isella, A., Huang, J., Andrews, S.~M., {et~al.} 2018, The Astrophysical
  Journal, 869, L49

\bibitem[{{Jermyn} \& {Kama}(2018)}]{jermyn2018}
{Jermyn}, A.~S. \& {Kama}, M. 2018, \mnras, 476, 4418

\bibitem[{{Kama} {et~al.}(2015){Kama}, {Folsom}, \& {Pinilla}}]{kama2015}
{Kama}, M., {Folsom}, C.~P., \& {Pinilla}, P. 2015, \aap, 582, L10

\bibitem[{{Kama} {et~al.}(2019){Kama}, {Shorttle}, {Jermyn}, {Folsom},
  {Furuya}, {Bergin}, {Walsh}, \& {Keller}}]{kama2019}
{Kama}, M., {Shorttle}, O., {Jermyn}, A.~S., {et~al.} 2019, \apj, 885, 114

\bibitem[{{Kamp} {et~al.}(2013){Kamp}, {Thi}, {Meeus}, {Woitke}, {Pinte},
  {Meijerink}, {Spaans}, {Pascucci}, {Aresu}, \& {Dent}}]{kamp2013}
{Kamp}, I., {Thi}, W.~F., {Meeus}, G., {et~al.} 2013, \aap, 559, A24

\bibitem[{{Kingdon} \& {Ferland}(1996)}]{kingdon1996}
{Kingdon}, J.~B. \& {Ferland}, G.~J. 1996, \apjs, 106, 205

\bibitem[{{Kingdon} \& {Ferland}(1999)}]{kingdon1999}
{Kingdon}, J.~B. \& {Ferland}, G.~J. 1999, \apj, 516, L107

\bibitem[{{Klarmann} {et~al.}(2018){Klarmann}, {Ormel}, \&
  {Dominik}}]{klarmann18}
{Klarmann}, L., {Ormel}, C.~W., \& {Dominik}, C. 2018, \aap, 618, L1

\bibitem[{{Krijt} {et~al.}(2018){Krijt}, {Schwarz}, {Bergin}, \&
  {Ciesla}}]{krijt18}
{Krijt}, S., {Schwarz}, K.~R., {Bergin}, E.~A., \& {Ciesla}, F.~J. 2018, \apj,
  864, 78

\bibitem[{{Lodders}(2010)}]{lodders2010}
{Lodders}, K. 2010, Astrophysics and Space Science Proceedings, 16, 379

\bibitem[{{Madhusudhan}(2012)}]{madhusudhan12b}
{Madhusudhan}, N. 2012, \apj, 758, 36

\bibitem[{{Mathurin} {et~al.}(2019){Mathurin}, {Dartois}, {Pino}, {Engrand},
  {Duprat}, {Deniset-Besseau}, {Borondics}, {Sandt}, \& {Dazzi}}]{mathurin19}
{Mathurin}, J., {Dartois}, E., {Pino}, T., {et~al.} 2019, \aap, 622, A160

\bibitem[{{McClure}(2019)}]{mcclure2019}
{McClure}, M.~K. 2019, arXiv e-prints, arXiv:1910.06029

\bibitem[{{McClure} {et~al.}(2013{\natexlab{a}}){McClure}, {Calvet},
  {Espaillat}, {Hartmann}, {Hern{\'a}ndez}, {Ingleby}, {Luhman}, {D'Alessio},
  \& {Sargent}}]{mcclure+13a}
{McClure}, M.~K., {Calvet}, N., {Espaillat}, C., {et~al.} 2013{\natexlab{a}},
  \apj, 769, 73

\bibitem[{{McClure} {et~al.}(2013{\natexlab{b}}){McClure}, {D'Alessio},
  {Calvet}, {Espaillat}, {Hartmann}, {Sargent}, {Watson}, {Ingleby}, \&
  {Hern{\'a}ndez}}]{mcclure+13b}
{McClure}, M.~K., {D'Alessio}, P., {Calvet}, N., {et~al.} 2013{\natexlab{b}},
  \apj, 775, 114

\bibitem[{{Menu} {et~al.}(2014){Menu}, {van Boekel}, {Henning}, {Chand ler},
  {Linz}, {Benisty}, {Lacour}, {Min}, {Waelkens}, {Andrews}, {Calvet},
  {Carpenter}, {Corder}, {Deller}, {Greaves}, {Harris}, {Isella}, {Kwon},
  {Lazio}, {Le Bouquin}, {M{\'e}nard}, {Mundy}, {P{\'e}rez}, {Ricci},
  {Sargent}, {Storm}, {Testi}, \& {Wilner}}]{menu14}
{Menu}, J., {van Boekel}, R., {Henning}, T., {et~al.} 2014, \aap, 564, A93

\bibitem[{{Owen} \& {Encrenaz}(2003)}]{owen03}
{Owen}, T. \& {Encrenaz}, T. 2003, \ssr, 106, 121

\bibitem[{{Piso} {et~al.}(2016){Piso}, {Pegues}, \& {{\"O}berg}}]{piso2016}
{Piso}, A.-M.~A., {Pegues}, J., \& {{\"O}berg}, K.~I. 2016, \apj, 833, 203

\bibitem[{{Rab} {et~al.}(2018){Rab}, {G{\"u}del}, {Woitke}, {Kamp}, {Thi},
  {Min}, {Aresu}, \& {Meijerink}}]{rab18}
{Rab}, C., {G{\"u}del}, M., {Woitke}, P., {et~al.} 2018, \aap, 609, A91

\bibitem[{{Rice} {et~al.}(2006){Rice}, {Armitage}, {Wood}, \&
  {Lodato}}]{rice2006}
{Rice}, W.~K.~M., {Armitage}, P.~J., {Wood}, K., \& {Lodato}, G. 2006, \mnras,
  373, 1619

\bibitem[{{Rubin} {et~al.}(2019){Rubin}, {Altwegg}, {Balsiger}, {Berthelier},
  {Combi}, {De Keyser}, {Drozdovskaya}, {Fiethe}, {Fuselier}, {Gasc},
  {Gombosi}, {H{\"a}nni}, {Hansen}, {Mall}, {R{\`e}me}, {Schroeder},
  {Schuhmann}, {S{\'e}mon}, {Waite}, {Wampfler}, \& {Wurz}}]{rubin19}
{Rubin}, M., {Altwegg}, K., {Balsiger}, H., {et~al.} 2019, \mnras, 489, 594

\bibitem[{{Schwarz} {et~al.}(2016){Schwarz}, {Bergin}, {Cleeves}, {Blake},
  {Zhang}, {{\"O}berg}, {van Dishoeck}, \& {Qi}}]{schwarz+16a}
{Schwarz}, K.~R., {Bergin}, E.~A., {Cleeves}, L.~I., {et~al.} 2016, \apj, 823,
  91

\bibitem[{{Shaw} {et~al.}(2005){Shaw}, {Ferland}, {Abel}, {Stancil}, \& {van
  Hoof}}]{shaw2005}
{Shaw}, G., {Ferland}, G.~J., {Abel}, N.~P., {Stancil}, P.~C., \& {van Hoof},
  P.~A.~M. 2005, \apj, 624, 794

\bibitem[{{Simcoe} {et~al.}(2013){Simcoe}, {Burgasser}, {Schechter}, {Fishner},
  {Bernstein}, {Bigelow}, {Pipher}, {Forrest}, {McMurtry}, {Smith}, \&
  {Bochanski}}]{simcoe13}
{Simcoe}, R.~A., {Burgasser}, A.~J., {Schechter}, P.~L., {et~al.} 2013, \pasp,
  125, 270

\bibitem[{{Sokal} {et~al.}(2018){Sokal}, {Deen}, {Mace}, {Lee}, {Oh}, {Kim},
  {Kidder}, \& {Jaffe}}]{sokal2018}
{Sokal}, K.~R., {Deen}, C.~P., {Mace}, G.~N., {et~al.} 2018, \apj, 853, 120

\bibitem[{{van Boekel} {et~al.}(2017){van Boekel}, {Henning}, {Menu}, {de
  Boer}, {Langlois}, {M{\"u}ller}, {Avenhaus}, {Boccaletti}, {Schmid},
  {Thalmann}, {Benisty}, {Dominik}, {Ginski}, {Girard}, {Gisler}, {Lobo Gomes},
  {Menard}, {Min}, {Pavlov}, {Pohl}, {Quanz}, {Rabou}, {Roelfsema}, {Sauvage},
  {Teague}, {Wildi}, \& {Zurlo}}]{vanboekel2017}
{van Boekel}, R., {Henning}, T., {Menu}, J., {et~al.} 2017, \apj, 837, 132

\bibitem[{{van 't Hoff} {et~al.}(2017){van 't Hoff}, {Walsh}, {Kama},
  {Facchini}, \& {van Dishoeck}}]{vanthoff2017}
{van 't Hoff}, M.~L.~R., {Walsh}, C., {Kama}, M., {Facchini}, S., \& {van
  Dishoeck}, E.~F. 2017, \aap, 599, A101

\bibitem[{{Vernet} {et~al.}(2011){Vernet}, {Dekker}, {D'Odorico}, {Kaper},
  {Kjaergaard}, {Hammer}, {Randich}, {Zerbi}, {Groot}, {Hjorth}, {Guinouard},
  {Navarro}, {Adolfse}, {Albers}, {Amans}, {Andersen}, {Andersen}, {Binetruy},
  {Bristow}, {Castillo}, {Chemla}, {Christensen}, {Conconi}, {Conzelmann},
  {Dam}, {de Caprio}, {de Ugarte Postigo}, {Delabre}, {di Marcantonio},
  {Downing}, {Elswijk}, {Finger}, {Fischer}, {Flores}, {Fran{\c{c}}ois},
  {Goldoni}, {Guglielmi}, {Haigron}, {Hanenburg}, {Hendriks}, {Horrobin},
  {Horville}, {Jessen}, {Kerber}, {Kern}, {Kiekebusch}, {Kleszcz}, {Klougart},
  {Kragt}, {Larsen}, {Lizon}, {Lucuix}, {Mainieri}, {Manuputy}, {Martayan},
  {Mason}, {Mazzoleni}, {Michaelsen}, {Modigliani}, {Moehler}, {M{\o}ller},
  {Norup S{\o}rensen}, {N{\o}rregaard}, {P{\'e}roux}, {Patat}, {Pena}, {Pragt},
  {Reinero}, {Rigal}, {Riva}, {Roelfsema}, {Royer}, {Sacco}, {Santin},
  {Schoenmaker}, {Spano}, {Sweers}, {Ter Horst}, {Tintori}, {Tromp}, {van
  Dael}, {van der Vliet}, {Venema}, {Vidali}, {Vinther}, {Vola}, {Winters},
  {Wistisen}, {Wulterkens}, \& {Zacchei}}]{vernet11}
{Vernet}, J., {Dekker}, H., {D'Odorico}, S., {et~al.} 2011, \aap, 536, A105

\bibitem[{{Zhang} {et~al.}(2017){Zhang}, {Bergin}, {Blake}, {Cleeves}, \&
  {Schwarz}}]{zhang2017}
{Zhang}, K., {Bergin}, E.~A., {Blake}, G.~A., {Cleeves}, L.~I., \& {Schwarz},
  K.~R. 2017, Nature Astronomy, 1, 0130

\bibitem[{{Zhang} {et~al.}(2019){Zhang}, {Bergin}, {Schwarz}, {Krijt}, \&
  {Ciesla}}]{zhang2019}
{Zhang}, K., {Bergin}, E.~A., {Schwarz}, K., {Krijt}, S., \& {Ciesla}, F. 2019,
  \apj, 883, 98

\bibitem[{{Zhang} {et~al.}(2020){Zhang}, {Bosman}, \& {Bergin}}]{zhang2020}
{Zhang}, K., {Bosman}, A.~D., \& {Bergin}, E.~A. 2020, \apjl, 891, L16

\bibitem[{{Zhang} {et~al.}(2013){Zhang}, {Pontoppidan}, {Salyk}, \&
  {Blake}}]{zhang13}
{Zhang}, K., {Pontoppidan}, K.~M., {Salyk}, C., \& {Blake}, G.~A. 2013, \apj,
  766, 82

\end{thebibliography}

%%%%%%%%%%%%%%%%%%%%%%%%%%%%%%%%%%%%%%%%%%%%%%%%%%%%%%%%%%%%%%%%%%%%%%%%%%
%%%%%%           APPENDIX #1              %%%%%%%%%%%%%%%%%%%%%%%%%%%%%%%%%%%%%%%%%%%%%%%%%%%%%
%%%%%%%%%%%%%%%%%%%%%%%%%%%%%%%%%%%%%%%%%%%%%%%%%%%%%%%%%%%%%%%%%%%%%%%%%%
\begin{appendix} %First appendix
\onecolumn
\section{Observation and data reduction details} %Second appendix
\label{obsdat}
One spectrum of TW Hya was obtained with the FIRE spectrograph \citep{simcoe13} at Las Campanas (R$\sim$6000, 0.8 < $\lambda$ < 2.5 $\mu$m)(30) on January 2nd, 2013 (PI McClure). Two 6.0 second exposures in the Fowler 1 read mode were obtained with the 0.6" slit under 0.45" seeing at airmass 1.009 in excellent conditions. We obtained the usual suite of arc lamp and flat-field calibrations and extracted the FIRE spectrum using the standard FIREhose pipeline, with MCN 7202 as the telluric calibrator star. The other two spectra were obtained with VLT X-shooter \citep{vernet11} (visible: 0.8 - 1 $\mu$m, R$\sim$18,000; near-IR arm: 1 - 2.5 $\mu$m, R$\sim$11,000) on April 7th and May 3rd, 2010. We corrected ESO's Phase 3 reduced data for telluric absorption using the MolecFit software package. Spectra from all three epochs were corrected for their heliocentric velocities, and the X-Shooter spectra were convolved down to the resolution of the FIRE spectrum. The processed spectra, before and after local continuum subtraction (as described in \S \ref{analysis}), are shown below. A comparable spectrum of hydrogen is shown in the main text (Fig. \ref{pagamma_fit}). The integrated line fluxes are given in Table \ref{tab_flux_abs}.

%---------- FIGURE ----------------
   \begin{figure*}
   \centering
   \includegraphics[width=17cm]{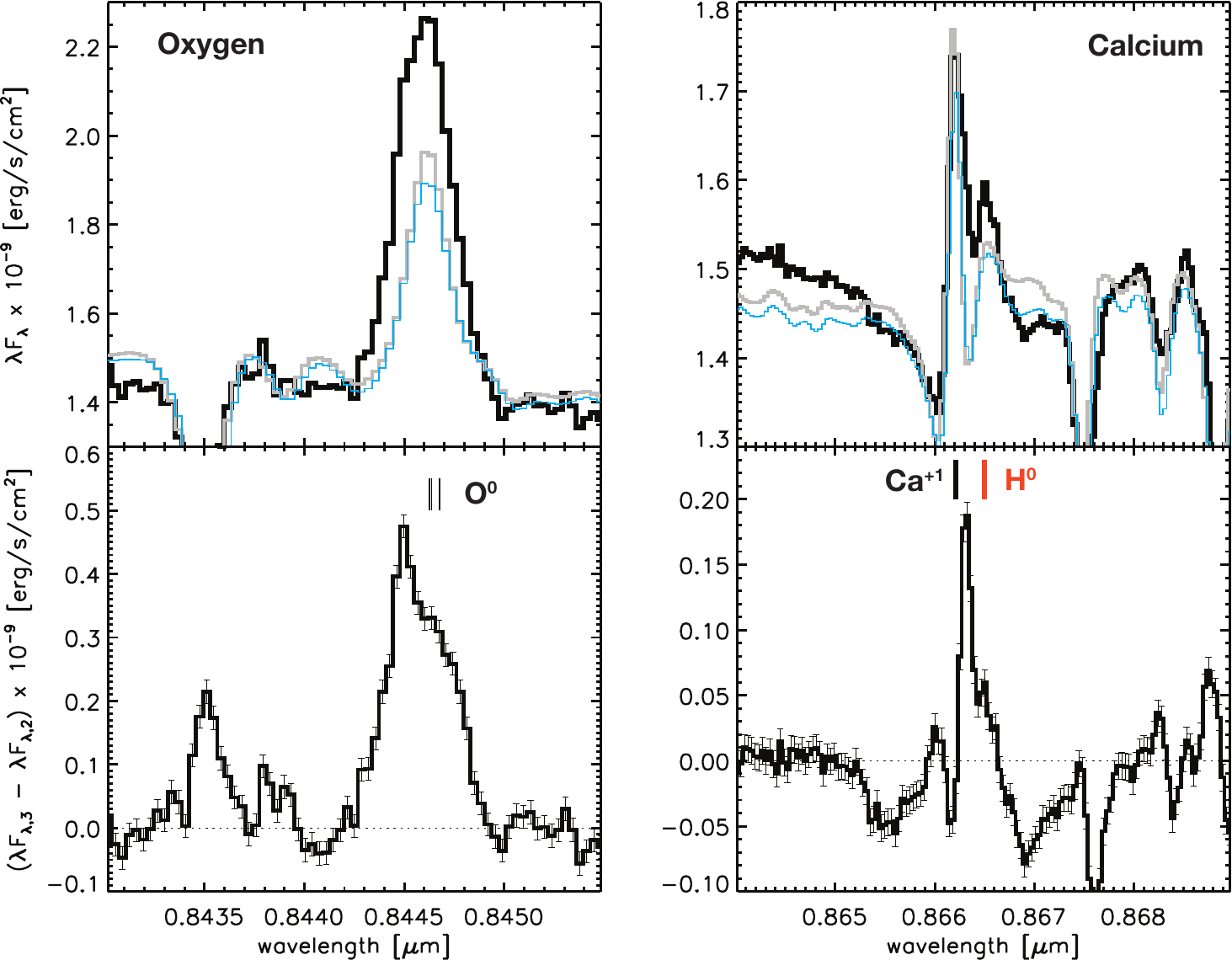}
   \caption{{\bf Top:} Comparison of the three observed epochs of TW Hya: 2010a (cyan, thin line), 2010b (gray, medium line), and 2013 (black, thick line). The oxygen line (left) is a triple line blend, and the calcium line (right) is blended with an H$^0$ Paschen series line and an absorption artifact from the telluric correction of that line, which we fit out before determining the integrated flux. {\bf Bottom:} Residual spectra after subtraction of epoch 2 from epoch 3.}
  \label{fluxes_1}%
              
  \end{figure*}
%-------------------------------------- 

%---------- FIGURE ----------------
   \begin{figure*}
   \centering
   \includegraphics[width=17cm]{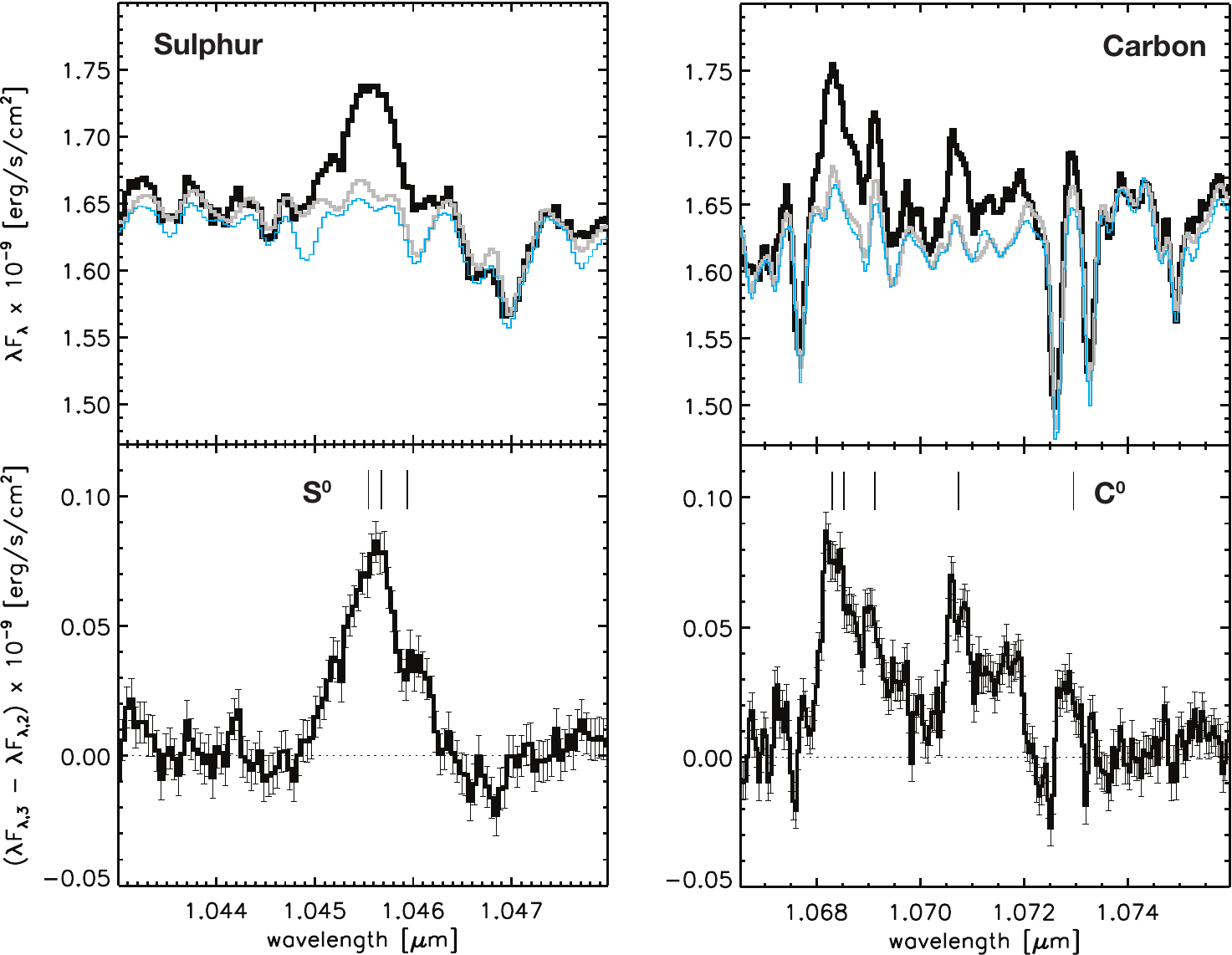}
   \caption{The same as for Fig. \ref{fluxes_1}, but for S (left) and C (right). The integrated flux of the S line is includes the three-line blend, while the C flux includes only the first three lines. Line styles are the same as in Fig. \ref{fluxes_1}.}
  \label{fluxes_2}%
              
  \end{figure*}
%-------------------------------------- 

%---------- FIGURE ----------------
   \begin{figure*}
   \centering
   \includegraphics[width=19cm]{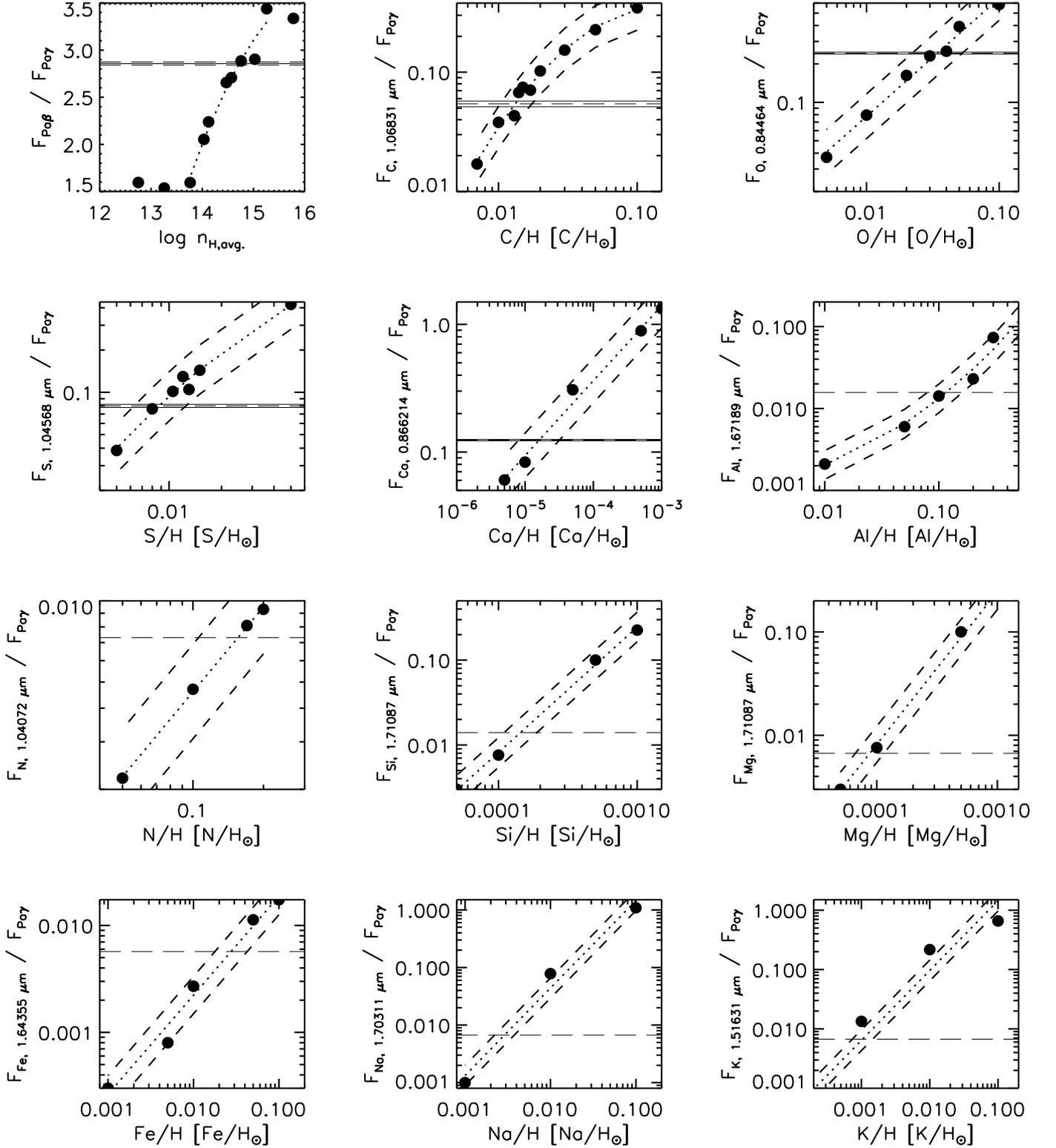}
   \caption{{\bf Upper left corner}: Model flux ratios between Pa $\beta$ and Pa $\gamma$ (y-axis), versus n$_{\rm H}$ (x-axis), compared with the observed flux ratio and error bars (horizontal lines). {\bf All other panels}: Model flux ratios of studied atomic lines relative to Pa$\gamma$ (y-axis) versus elemental abundance relative to H in solar units (x-axis) for all elements in this study. The diagonal dashed lines indication the systematic error in the models due to the charge exchange rate coefficients, as discussed in Appendix \ref{modapp}. Observational flux ratios are indicated (horizontal lines), with error bars for the four elements with $>3\sigma$ detections (C, O, S, and Ca), while 3$\sigma$ upper limits are indicated for the other elements. }
  \label{fluxes_3}%
              
  \end{figure*}
%-------------------------------------- 

\begin{table*}
\caption{Observed line fluxes and modeled abundances}             % title of Table
\label{table:1}      % is used to refer this table in the text
\centering                          % used for centering table
\begin{tabular}{c c c c c c}        % centered columns (4 columns)
\hline\hline                 % inserts double horizontal lines
 \\
Atomic species			&	Line  	&	F$_{int}$ 				&	S/N  		&	F$_X$/F$_{Pa \gamma}$ 	&	X/H [(X/H)$_{\odot}$]	\\
					&  [$\mu$m]	&	[erg/s/cm$^{-2}$]		&			&						&						\\
\hline \\
H$^0$ (Pa $\gamma$)	&	1.09381	&	7.0$\times$10$^{-9}$	&	187		&	-				&		\\
H$^0$ (Pa $\beta$)		&	1.28181	&	2.0$\times$10$^{-8}$	&	1090.5	&	2.857$\pm$0.016	&	log (n$_{\rm H}$) = 14.74$\pm$0.02	\\
\hline \\
C$^0$				&	1.06831	&	3.8$\times$10$^{-10}$	&	18.3		&	0.0543$\pm$0.0030	&	1.26$\pm0.54$ $\times$10$^{-2}$	\\
N$^0$ (blend)			&	1.04072	&	$<$2.5$\times$10$^{-10}$	&	< 3		&	$<$0.0073		&	$<$ 2.35$\times$10$^{-1}$	 \\
O$^0$ (blend)			&	0.84464	&	1.7$\times$10$^{-9}$	&	58.7		&	0.2429$\pm$0.0043	&	3.27$\pm$2.09 $\times$10$^{-2}$	\\
Na$^0$				&	1.70311	&	$<$4.7$\times$10$^{-11}$	&	< 3		&	$<$0.0067		&	$<$2.67$\times$10$^{-3}$	\\
Mg$^0$				&	1.71087	&	$<$4.7$\times$10$^{-11}$	&	< 3		&	$<$0.0067		&	$<$1.12$\times$10$^{-4}$	\\
Al$^0$				&	1.67189	&	$<$1.1$\times$10$^{-10}$	&	< 3		&	$<$0.0157		&	$<$1.68$\times$10$^{-1}$ 	\\
Si$^0$				&	1.58884	&	$<$9.8$\times$10$^{-11}$	&	< 3		&	$<$0.0140		&	$<$1.83$\times$10$^{-4}$ 	\\
S$^0$ (blend)			&	1.04568	&	5.6$\times$10$^{-10}$	&	39.8		&	0.0800$\pm$0.0021	&	8.70$\pm4.06$ $\times$10$^{-3}$ \\
K$^0$				&	1.51631	&	$<$4.7$\times$10$^{-11}$	&	< 3		&	$<$0.0067		&	$<$ 6.33$\times$10$^{-3}$	\\
Ca$^{+1}$				&	0.86621	&	8.7$\times$10$^{-10}$	&	93.1		&	0.1240$\pm$0.0015	&	1.39$\pm$1.15$\times$10$^{-5}$ \\
Fe$^{+1}$				&	1.64355	&	$<$4.0$\times$10$^{-11}$	&	< 3		&	$<$0.0057		&	$<$ 2.72$\times$10$^{-2}$	\\
\hline         \\                          %inserts single line
\end{tabular}
 \\
 \label{tab_flux_abs}
 Notes: The uncertainties on the modeled abundances are dominated by a systematic 20\% uncertainty on the synthetic line fluxes, which results from factor of $\sim$2-4 uncertainty on the charge exchange rate coefficients, as discussed in Section 3.3.4 of Hazy, the Cloudy users' manual. Abundances are with respect to the \citet{asplund2009} solar values.
\end{table*}

%--------------------------------------------------------------------------------------------------------------------------------------------------------------------------------------
\section{Inner disk chemical model mechanisms and uncertainties} %Second appendix
\label{modapp}

Our 1D inner disk model uses version 17.02 of Cloudy \citep{ferland2017}. As discussed in Section 3.3 of \citet{mcclure2019}, the high midplane gas densities at 0.024 AU (n$_{\rm H}\sim$10$^{15}$ [cm$^{-3}$], N$_{\rm H}\sim10^{26}$ [cm$^{-2}$]) lead to a chain of reactions that ultimately produce a mostly neutral gas population, but with a large fraction of free electrons donated mainly by hydrogen itself. 

\begin{figure*}
   \centering
   \includegraphics[width=8cm,angle=90]{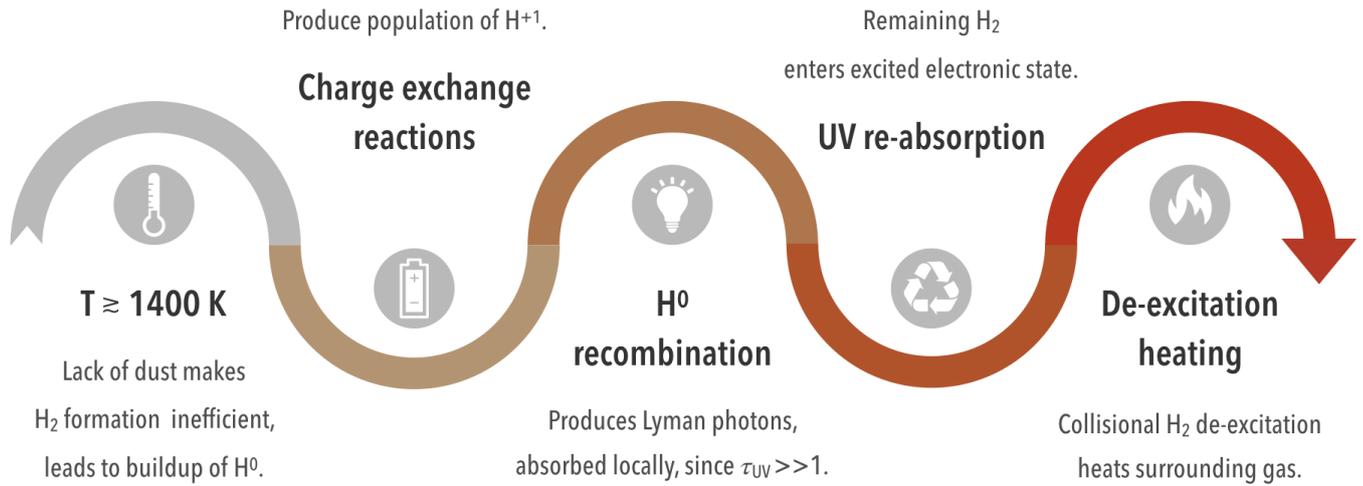}
   \caption{{\bf Flowchart of the inner disk physical processes} leading to the high temperatures and domination of H over H$_2$ and C over CO at the disk midplane, in contrast to the typical photon-dominated region physics seen in molecular clouds and the outer disk upper layers.}
  \label{figb1}%
              
  \end{figure*}

Since the physics and chemistry of molecular hydrogen is critical to understand the atomic H emission, we enabled the larger H$_2$ molecule module in Cloudy \citep{shaw2005}. Collisional dissociation of H$_2$ is efficient at these high densities; however, formation of H$_2$ is hampered, due to the absence of dust inside the silicate sublimation rim (at 0.3 AU). Therefore a large population of H$^0$ exists. In turn, at these densities H$^{+1}$ is efficiently created through charge exchange reactions, particularly with sulphur as the collision partner. The fraction of hydrogen in H$^{+1}$ is $\sim$10$^{-3.5}$, and therefore H is the major contributor to the electron density, unlike a PDR where C$^{+1}$ dominates. Recombination to H$^0$ produces Lyman photons. However, at these large column densities the gas is optically thick to UV radiation, so the Lyman photons are absorbed locally by the remaining population of H$_2$, causing it to enter an excited electronic state. Collisional de-excitation of the H$_2$ then heats the surrounding gas to $\sim$6000 K. The net result is that at this location in the disk midplane there is no dust, little molecular emission, and mostly neutral atomic gas for which emission emission line ratios directly probe the bulk elemental abundances. We visually summarize this series of processes in Fig. \ref{figb1}.

The main source of uncertainty in these calculations comes from the charge exchange rate coefficients, which are accurate to within a factor of 2-4 \citep{kingdon1996, kingdon1999}. The coefficients determine the amount of H$^{+1}$, and therefore the electron density of the gas. In turn, the electron density impacts the level populations in the upper state of the transitions giving rise to the observed emission lines. The uncertainty in the rate coefficients leads to a systematic uncertainty of typically 20\% on the synthetic line fluxes, as discussed in Section 3.3.4 of Hazy, the Cloudy users' manual. We have included that uncertainty in the abundances listed in Table \ref{tab_flux_abs}. A secondary source of uncertainty in these calculations is the geometry of the emitting region. However, because we are fitting line ratios, we are not overly sensitive to the geometry of the emitting area, assuming that all of the lines originate in the same region.

An additional complication in the N abundance determination in less dense regions would normally be self-shielding of N$_2$ against photodissociation, which is known to limit conversion of N$_2$ into atomic N or other N-bearing molecules more efficiently than for CO. However, as the excitation temperature increases, the efficiency of self-shielding relative to other modes of shielding (e.g. by H, H$_2$, or CO) becomes much less \citep{heays2014}, and a larger column of N$_2$ would be required in order to self-shield. At the high temperatures and densities in our model (T$_e\sim$6500 K, log (n$_H$)$\sim$14.75 cm$^{-3}$), Cloudy indeed predicts that atomic N represents at least 99.8\% of the total N budget. Critically, even if shielding were efficient at these temperatures, the fact that the UV field is generated locally by H recombination would effectively prevent the formation of a sufficiently large column to shield, as the H+ and N$_2$ are intimately mixed. There is observational evidence, in any event, for N$_2$ photodissociation already occurring further out in the disk. \citet{hily-blant2019} detect selective N$_2$ photodissociation indirectly where the N$_2$ snowline starts in the disk upper layers around 40 AU, through variation in the $^{15}$N/$^{14}$N ratio in ALMA observations of HCN isotopologs. And in our own data, we see the 1.04$\mu$m [N I] quadruplet at all three epochs; however, it subtracts out of the epoch differenced spectrum, suggesting that it originates further out in the disk rather than in the ring of denser material that we model moving towards the star in 2013.

As an independent check on the resulting hydrogen density, we ran CHIANTI models of line ratios taken between the C$^0$ lines, as in \citet{mcclure2019}, with additional line ratios between the S$^0$ triplet. The range of electron densities and electron temperatures produced by these abundance- and geometry-independent measures for both C and S, combined, are log (n$_e$) $\sim$ 10.3-12.0 cm$^{-3}$ at T$_e$ $\sim$ 7500-5500 K. The best-fitting Cloudy model has a value of log (n$_e$) $\sim$ 11 cm$^{-3}$ at T$_e$ $\sim$ 6500 K, which is consistent with CHIANTI's predicted range.

 \section{Carbon locking fractions} %Second appendix
\label{c_lock}

As an independent check on the location where carbon is locked out of the accreting gas, we formulate a toy model, pictured in Figure \ref{fig1} of the main text.  First, we divide the disk into three zones: Zone 1 (between the outer disk edge and the CO snowline), Zone 2 (between the CO snowline and dust inner rim), and Zone 3 (interior to the dust inner rim). Then we define the model as a parcel of constant density gas and dust that initially sits in the outer disk but is radially transported via accretion processes through the disk to the launching region for the stellar magnetospheric accretion columns. The parcel starts as carbon gas, $M_{g,0}$, with overall solar carbon abundances $X_{g,0}$=$\big(C/H\big)_{\odot}$=2.69$\times10^{-4}$ \citep{asplund2009}, from which in Zone 1 solid carbon condenses out with mass: 
\begin{equation}
M_{d,1} = M_{g,0} - M_{g,1},
\end{equation}
\noindent leaving a depleted carbon gas mass of $M_{g,1}$ and corresponding abundance $X_{g,1}$ with respect to hydrogen. In the main text, we assume $X_{g,1}$=6.4$\times10^{-7}$ \citep{zhang2019}.

The solid carbon component can be split into volatile ice, as CO, and dust that sublimates at higher temperatures than CO, with fraction $f_r$. The higher temperature carbon could be in any of the following materials: CO$_2$ or other C-rich ices, organic residues, amorphous carbon, or graphite. The grains are assumed to move radially along with the gas, unless they become decoupled from the accretion stream and locked in place in one of the zones. This locking could be caused either by the dust decoupling from the gas and entrapment of larger, millimeter-sized dust grains at pressure maxima or by the formation of bodies that are large enough to neither accrete with the gas nor radially drift. The model is agnostic to the exact locking mechanism. We do not consider complications from decoupling of the gas and non-locked dust due to radial drift; we assume that this dust moves with the gas. 

We define a fraction, $f_{L,1}$, of the initial dust, $M_{d,1}$, that is locked in some manner in Zone 1, outside of the CO snowline, while the remaining dust exists in small, coupled grains that accrete with the gas across the CO snowline. The successfully transported dust is defined as: 
\begin{equation}
M_{d,d,1}=(1-f_{L,1})M_{d,1}. 
\end{equation}
\noindent At the CO snowline, the volatile dust sublimates, releasing carbon back into the gas phase. Therefore the new carbon gas mass is: 
\begin{equation}
M_{g,2}=M_{g,1}+(1-f_r)M_{d,d,1},
\end{equation}
\noindent which can be rewritten as: 
\begin{equation}
M_{g,2}=M_{g,1}+(1-f_r)(1-f_{L,1})(M_{g,0}-M_{g,1}).  \\
\label{mass2}
\end{equation}

\noindent and corresponds to a carbon abundance of $X_{g,2}$ with respect to hydrogen. In the main text, we assume $X_{g,2}=2.5\times10^{-6}$ \citep{zhang2019}. Solving Eq. \ref{mass2} for the fraction of dust locked in Zone 1, $f_{L,1}$, and rewriting the gas masses of each zone in terms of their observed C/H abundances, $X_{g,1}$ and $X_{g,2}$, yields: 
\begin{equation}
f_{L,1}=1-\frac{1}{1-f_r}\cdot\frac{X_{g,2}-X_{g,1}}{X_{g,0}-X_{g,1}} \\
\label{fl1}
\end{equation}
\noindent assuming that no appreciable hydrogen mass is locked into the grains. The dust mass interior to the CO snowline is then the fraction of the transported solids with a higher sublimation temperature than CO: 
\begin{equation}
M_{d,2}=f_rM_{d,d,1}.
\end{equation} 
\noindent A fraction of the dust mass in Zone 2, $f_{L,2}$, is also locked eventually, either through growth to large grains or to planetesimal sizes. Therefore, the mass of dust that is transported in the parcel across the inner dust rim is the dust mass in Zone 2 less the fraction of dust mass M$_{d,2}$ locked in Zone 2:
\begin{equation}
M_{d,d,2}=(1-f_{L,2})M_{d,2}.
\end{equation}
\noindent The mass of gas phase carbon interior to the innermost dust rim is then the carbon gas mass of Zone 2, plus the additional mass of higher temperature transported dust, all of which now sublimates:
\begin{equation}
M_{g,3}=M_{g,2}+M_{d,d,2}.
\end{equation}
\noindent The latter can be rewritten as:
\begin{equation}
M_{g,3}=M_{g,2}+(1-f_{L,2})f_r(1-f_{L,1})(M_{g,0}-M_{g,1}).
\label{mass3}
\end{equation}

\noindent The Zone 3 corresponds to the carbon abundance found in this work, $X_{g,3}=3.4\times10^{-6}$. Solving Eq. \ref{mass3} for the fraction of dust mass M$_{d,2}$ locked into Zone 2, $f_{L,2}$, yields:
\begin{equation}
f_{L,2}=1-\frac{1}{(1-f_{L,1})f_r}\cdot\frac{X_{g,3}-X_{g,2}}{X_{g,0}-X_{g,1}} \\
\label{fl2}
\end{equation}

\noindent This locked fraction, which is in terms of the Zone 2 dust mass, can be converted via the definition of M$_{d,2}$ into a locking fraction in terms of the initial dust mass, $M_{d,1}$. We cannot simultaneously constrain both the locking fraction and the higher temperature carbon fraction, but if we assume that the latter is equivalent to the refractory carbon fraction, then reasonable value range from $f_r=0.5$ in the ISM to $f_r=0.83$ in comet 67P, as discussed in the main text. Using this range of values and the three observed carbon abundances discussed above produces locking fractions exterior (Zone 1) and interior (Zone 2) to the CO snowline ranging from 90-99\%.

\end{appendix}


\end{document}